\documentclass[12pt]{article}
\usepackage{epsfig} \usepackage{amssymb} \usepackage{amsfonts}
\usepackage{psfig}

\newcommand{\bea}{\begin{eqnarray}}
\newcommand{\eea}{\end{eqnarray}}

\newcommand{\as}{\alpha_s}
\newcommand{\asMZ}{\alpha_s(M^2_Z)}

\begin{document}

%                          Title
\begin{center}
{\Large \bf A study of QCD coupling constant from \\
fixed target deep inelastic measurements
%Title of the paper
} \\

\vspace{4mm}

%                      author/address
%Author's name\\
V.G. Krivokhijine and A.V. Kotikov\\ 

\vspace{4mm}
Joint Institute for Nuclear
Research, 141980 Dubna, Russia
\end{center}

%                        Abstract
\begin{abstract}
We reanalyze 
%high statistic 
deep inelastic scattering data 
of BCDMS Collaboration by including proper cuts of  ranges
with large systematic errors. 
We perform also the 
%combine 
fits of high statistic deep inelastic scattering data 
of BCDMS, SLAC, NM and BFP Collaborations  
taking the data separately and in combined way and find good agreement
between these analyses. We
%and 
extract the values of 
the QCD coupling constant $\alpha_s(M^2_Z)$ up to
NLO level. 
The fits of the combined data for
the nonsinglet part of the structure function $F_2$
predict the coupling constant value $\alpha_s(M^2_Z) = 0.1174 \pm 0.0007$
(stat) $\pm 0.0019$ (syst) $\pm 0.0010$ (normalization). 
The fits of the combined data for
both: the nonsinglet part of $F_2$ and the singlet one, lead to the
%little higher 
values $\alpha_s(M^2_Z) = 0.1177 \pm 0.0007$
(stat) $\pm 0.0021$ (syst) $\pm 0.0009$ (normalization). 
Both above values are  in very good
 agreement with each
%one to 
other.
\end{abstract}

%\vspace{-4mm}
%\vskip -0.4cm

\section{ Introduction }

The deep inelastic scattering (DIS) leptons on hadrons is the basical
 process to study the values of the parton distribution functions (PDF)
which are universal (after choosing of factorization and renormalization 
schemes) and
can be used in other processes.
The accuracy of the present data for deep inelastic
%(DIS) 
structure functions (SF) reached the level at which
the $Q^2$-dependence of logarithmic QCD-motivated and power-like ones
%are observed and 
may be studied separately 
%(see, for example, the recent reviews 
(for a review, see the recent papers \cite{Beneke} and references 
therein).

In the present paper we review the results of our analysis \cite{KriKo}
at the next-to-leading (NLO)
 order \footnote{The evaluation
of $\alpha_s^3(Q^2)$ corrections to anomalous dimensions of Wilson operators,
that will be done in nearest future by Vermaseren and his coauthors (see 
discussions in \cite{Stirling}), 
gives a possibility
to apply many modern programs to perform fits of data at next-next-to-leading
order (NNLO) of perturbative theory (see detail discussions in 
%Section 5 Summary
\cite{KriKo}).}
%large experience in fits of data
of perturbative QCD for
the most known DIS SF 
$F_2(x,Q^2)$ taking into account SLAC, NMC,  BCDMS
and BFP
experimental data \cite{SLAC1}-\cite{BFP}.
We
stress the power-like effects, so-called twist-4 (i.e.
$\sim 1/Q^2$) 
%and twist-6 (i.e. $\sim 1/Q^4$) 
contributions.
To our purposes we represent the SF $F_2(x,Q^2)$ as the contribution
of the leading twist part $F_2^{pQCD}(x,Q^2)$ 
described by perturbative QCD and the  
nonperturbative part (i.e. twist-four terms $\sim 1/Q^2$):
\begin{equation}
F_2(x,Q^2) 
\equiv F_2^{full}(x,Q^2)
=F_2^{pQCD}(x,Q^2)\,\left(1+\frac{\tilde h_4(x)}{Q^2}\right) 
%+ \frac{\tilde h_6(x,Q^2)}{Q^4}
\label{1}
\end{equation}

The SF $F_2^{pQCD}(x,Q^2)$ obeys the (leading twist)
perturbative QCD dynamics
including the target mass corrections 
(and coincides with $F_2^{tw2}(x,Q^2)$
when the target mass corrections are withdrawn).

The Eq.(\ref{1}) allows us to separate pure kinematical power corrections,
i.e. so-called target mass corrections, so that the function 
$\tilde h_4(x)$ corresponds to ``dynamical''
contribution of the twist-four operators. The parameterization (\ref{1})
implies 
\footnote{The r.h.s. of the Eq.(\ref{1}) is represented sometimes
as  $F_2^{pQCD}(x,Q^2)+\overline h_4(x)/Q^2$. It implies that 
the anomalous dimensions of the twist-four operators 
are equal to zero.} 
that the
anomalous dimensions of the twist-two and twist-four operators are equal
to each other, that is not correct in principle.  Meanwhile,
in view of limited precision of the data, the approximation (\ref{1})
and one in the footnote 2 give rather good predictions (see discussions in 
\cite{Al2000}).

Contrary to standard fits (see, for example, \cite{Al2000}- \cite{fits}) 
when the direct
numerical calculations based on Dokshitzer-Gribov-Lipatov-Altarelli-Parisi
(DGLAP) equation \cite{DGLAP} are used to evaluate structure functions, 
we use the exact solution of DGLAP equation
for the Mellin moments $M_n^{tw2}(Q^2)$ of
%$F_2^{full}(x,Q^2)$, $F_2^{pQCD}(x,Q^2)$ and
SF $F_2^{tw2}(x,Q^2)$:
\begin{equation}
M_n^{k}(Q^2)=\int_0^1 x^{n-2}\,F_2^{k}(x,Q^2)\,dx~~~~~~~ (\mbox{hereafter }
k=full, pQCD, tw2, ...)
\label{2}
\end{equation}
and
the subsequent reproduction of $F_2^{full}(x,Q^2)$, $F_2^{pQCD}(x,Q^2)$  
and/or
$F_2^{tw2}(x,Q^2)$ at every needed $Q^2$-value with help of the Jacobi 
Polynomial expansion method \cite{Barker,Kri}
(see also similar analyses at the NLO level 
\cite{Vovk}
and at the NNLO level and above \cite{PKK}).

In this paper we consider in detail only fits of the nonsinglet part (NS)
of SF $F_2$
and review only final results in general case, that
%. The full analysis
can be found in \cite{KriKo}.
Moreover, we do not present exact formulae of $Q^2$-dependence
of SF $F_2$ which are also given in \cite{KriKo}. We note only that
the moments $M_{NS}(n,Q^2)$ at 
some $Q^2_0$ is theoretical input of our analysis and 
the twist-four term $\tilde h_4(x)$
is considered as a set free parameters at each $x_i$ bin. The set has the form
$\tilde h_4^{free}(x)=\sum_{i=1}^{I} \tilde h_4(x_i)$, 
where $I$ is the number of bins.
The
constants $\tilde h_4(x_i)$ (one per $x$-bin) parameterize $x$-dependence of 
$\tilde h_4^{free}(x)$.

%%%%%%%%%%%%%%%%  3  3  3  3 %%%%%%%%%%%%%%%%%%%

\vspace{-0.3cm}
\section{ Fits of $F_2$: procedure }

Having the QCD expressions for the Mellin moments
$M_n^{k}(Q^2)$ we can reconstruct the SF $F_2^k(x,Q^2)$
%using the Jacobi polynomial expansion method:
as
\vskip -0.3cm
\begin{equation}
F_{2}^{k,N_{max}}(x,Q^2)=x^{a}(1-x)^{b}\sum_{n=0}^{N_{max}}
\Theta_n ^{a , b}
(x)\sum_{j=0}^{n}c_{j}^{(n)}{(\alpha ,\beta )}
M_{j+2}^{k} \left ( Q^{2}\right ),
\label{2.1}
\end{equation}
%\vskip -0.3cm
where $\Theta_n^{a,b}$ are the Jacobi polynomials
\footnote{We would like to note here that there is similar method 
\cite{Ynd} to reproduce of structure functions, based on Bernstein 
polynomials. The method has been used 
in several analyses at the NLO level in \cite{KaKoYaF}
and at the NNLO level in \cite{SaYnd}.}
and $a,b$ are the parameters, fitted by the condition
of the requirement of the minimization of the error of the
reconstruction of the
structure functions  (see Ref.\cite{Kri} for details).

First of all, we choose the cut $Q^2 \geq 1$ GeV$^2$ in all our studies.
For $Q^2 < 1$ GeV$^2$, the applicability of twist expansion is very
questionable. 

Secondly, we
choose quite large values of the normalization point
$Q^2_0$. There are several reasons of this choice:
\begin{itemize}
\item 
Our perturbative formulae should be applicable at the value of
$Q^2_0$. Moreover, the higher order corrections $\sim \as^n(Q^2_0)$ 
($n \geq 2$), coming from normalization conditions of PDF,
 are less important at higher $Q^2_0$ values.
\item
It is necessary to cross heavy quark thresholds less number of time
to reach $Q^2=M^2_Z$, the point of QCD coupling constant normalization.
\item
It is better to have the value of $Q^2_0$ around the middle point
of {\it logarithmical} range of considered $Q^2$ values. 
%At least,
Then
at the case the higher order corrections $\sim (\as(Q^2)-\as(Q^2_0))^n$ 
($n \geq 2$) are less important.
\end{itemize}

%\vspace{0.5cm}

We use MINUIT program \cite{MINUIT} for
minimization of two $\chi^2 $ values:
\begin{eqnarray}
\chi^2(F_2) = {\biggl|\frac{F_2^{exp} - F_2^{teor}}{\Delta F_2^{exp}}
\biggr| }^2 
\nonumber
\end{eqnarray}
%where $b=dlnF_2/dlnQ^2$.

We would like to apply the following procedure: we study the dependence of 
$\chi^2/DOF$ value on value of $Q^2$ cuts for various sets of experimental
data. The study will be done for the both cases: including higher twists
%terms 
corrections  and without them.

We use free normalizations of data for different experiments. 
For the reference, we use the most stable deuterium BCDMS data
at the value of energy $E_0=200$ GeV 
\footnote{$E_0$ is the initial energy lepton beam.}. 
%The usage 
Using other types of data as reference gives
negligible changes in our results. The usage of fixed normalization
for all data leads to fits with a bit worser $\chi^2$.

\vspace{-0.3cm}
\section{ Results of fits }

Hereafter 
%at nonsinglet case of evolution 
we choose
 $Q^2_0$ = 90 GeV$^2$ that is in good agreement with above 
conditions. We use also $N_{max} =8$,
the cut $0.25 \leq x \leq 0.8$, where the nonsinglet evolution is
dominant.

\vspace{-0.2cm}
\subsection { BCDMS $C^{12} + H_2 + D_2$ data }
%{\bf 4.1.} {\bf SLAC data}

We start our analysis with the most precise experimental data 
\cite{BCDMS1} obtained  by BCDMS muon
%$\mu h$ 
scattering experiment at the high $Q^2$ values.
The full set of data is 607 points.

It is well known that the original analyses 
%of 
given by BCDMS Collaboration itself (see
also Ref. \cite{ViMi}) lead to quite small values 
 $\alpha_s(M^2_Z)=0.113$.
Although in some recent papers (see, for example, 
\cite{Al2000,H1BCDMS})
more higher values of $\alpha_s(M^2_Z)$ have been observed, we think that
an additional reanalysis of BCDMS data should be very useful. 

Based on study \cite{Kri2} (see also \cite{H1BCDMS}) we proposed in
\cite{KriKo} that 
the reason for small values
of $\alpha_s(M^2_Z)$ coming from BCDMS data was the existence of the subset
of the data having large systematic errors. 
We studied this subject by 
introducing several so-called $Y$-cuts 
\footnote{Hereafter we use the kinematical variable $Y=(E_0-E)/E_0$,
where $E_0$ and $E$ are initial and scattering energies of lepton, 
respectively.}
(see \cite{Kri2}). Excluding this set of data with large systematic errors
leads to essentially larger values of $\alpha_s(M^2_Z)$ and very slow
dependence of the values on the concrete choice of the $Y$-cut (see below).

We studied influence of the experimental systematic errors on the
results of the QCD analysis as a function of $Y_{cut3}$, $Y_{cut4}$ and
$Y_{cut5}$ applied to the data.
We use the following $x$-dependent $y$-cuts:
\bea
& &y \geq 0.14 \,~~~\mbox{ when }~~~ 0.3 < x \leq 0.4 \nonumber \\
& &y \geq 0.16 \,~~~\mbox{ when }~~~ 0.4 < x \leq 0.5 \nonumber \\
& &y \geq Y_{cut3} ~~~\mbox{ when }~~~ 0.5 < x \leq 0.6 \nonumber \\
& &y \geq Y_{cut4} ~~~\mbox{ when }~~~ 0.6 < x \leq 0.7 \nonumber \\
& &y \geq Y_{cut5} ~~~\mbox{ when }~~~ 0.7 < x \leq 0.8 
\label{cut}
\eea
and
%We use 
several sets $N$ of the values for the cuts at $0.5 < x \leq 0.8$
%, which are 
given in the Table.

\begin{center}
%\footnotesize
\begin{tabular}{|c|c|c|c|c|c|c|c|}
\hline
%& & & & & & & \\
$N$ & 0 & 1 & 2 & 3 & 4 & 5 & 6 \\
%& & & & & & & \\
\hline \hline
$Y_{cut3}$ & 0 & 0.14 & 0.16 & 0.16 & 0.18 & 0.22 & 0.23 \\  
%\hline
$Y_{cut4}$ & 0 & 0.16 & 0.18 & 0.20 & 0.20 & 0.23 & 0.24 \\
$Y_{cut5}$ & 0 & 0.20 & 0.20 & 0.22 & 0.22 & 0.24 & 0.25 \\
\hline
\end{tabular}
\end{center}

%\vspace{0.2cm}
{\bf Table. } The values of $Y_{cut3}$, $Y_{cut4}$ and $Y_{cut5}$.
\vspace{0.4cm}

The systematic errors for BCDMS data are given \cite{BCDMS1}
as multiplicative factors to be applied to $F_2(x,Q^2)$: $f_r, f_b, f_s, f_d$
and $f_h$ are the uncertainties due to spectrometer resolution, beam momentum,
calibration, spectrometer magnetic field calibration, detector inefficiencies 
and energy normalization, respectively.

For this study each experimental point of the undistorted set was multiplied
by a factor characterizing a given type
of uncertainties and a new (distorted) data set was fitted again
in agreement with our procedure considered in the previous section. The factors
($f_r, f_b, f_s, f_d, f_h$) were taken from papers \cite{BCDMS1}
(see CERN preprint versions in \cite{BCDMS1}).
The absolute differences between the values of $\alpha_s$ for the distorted
and undistorted sets of data are given in 
%Table 2 and 
the Fig. 1 as the total systematic
error of $\alpha_s$ estimated in quadratures. The number of the experimental
points and the value of $\alpha_s$ for the undistorted set of $F_2$ are also
presented in 
%the Table 2 and 
the Fig. 1.

%\newpage

   \begin{figure}[tb]\label{fig-1}
\unitlength=1mm
\vskip -1.5cm
\begin{picture}(0,100)
  \put(0,-5){%
   \psfig{file=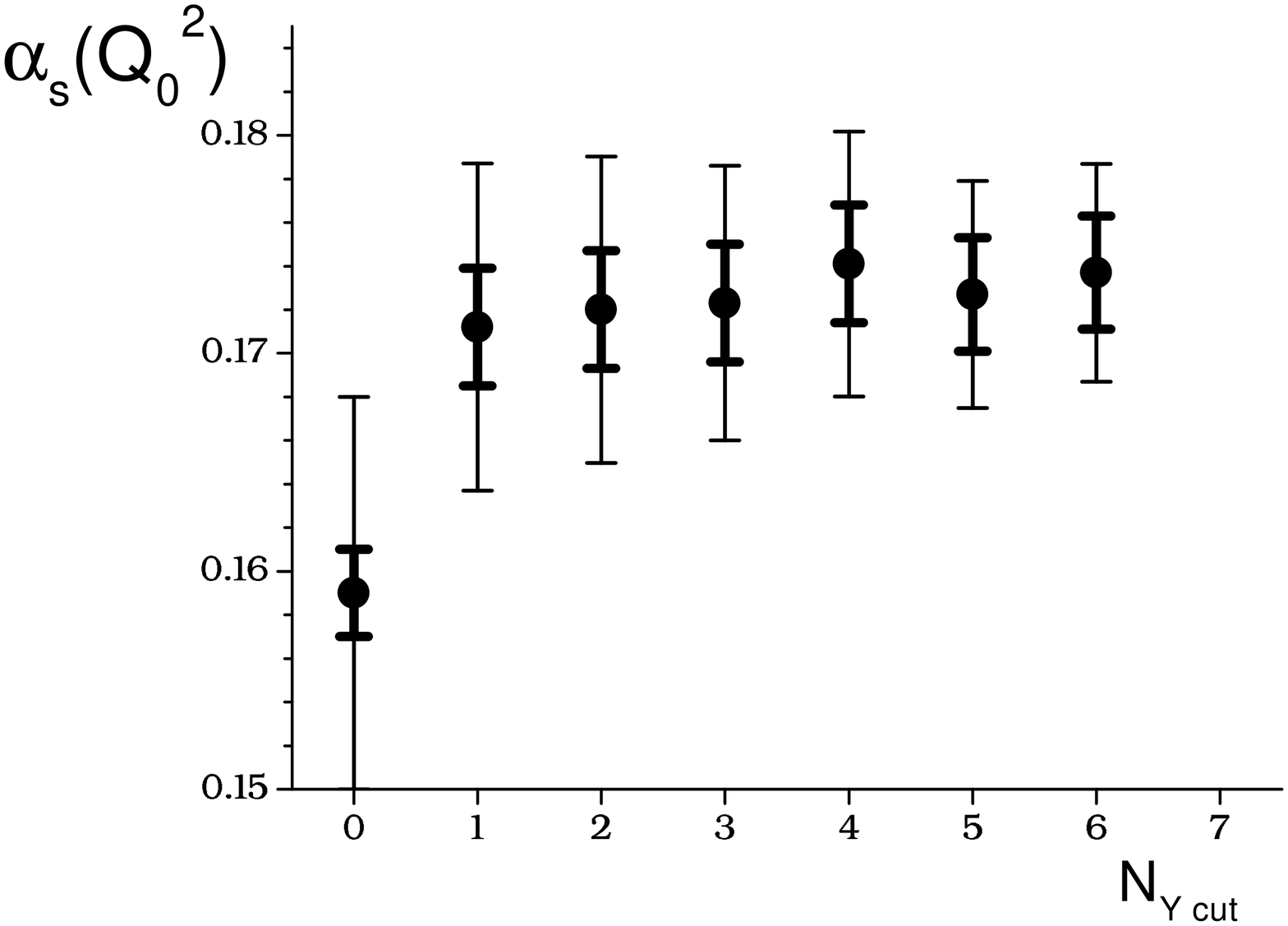,width=170mm,height=90mm}%
}
\end{picture}
\vskip 0cm
 \caption{
The study of systimatics at different $Y_{cut}$ values.
%in the fits based on nonsinglet evolution.
The  QCD analysis of BCDMS $C^{12}, H_2, D_2$ data:
the inner (outer) error-bars show statistical (systematic) errors.
}
\vskip 0cm
 \end{figure}
%\end{document}

From 
%the Table 2 and 
the Fig. 1 we can see that the $\alpha_s$ values are obtained
for $N=1 \div 6$ of $Y_{cut3}$, $Y_{cut4}$ and $Y_{cut5}$ are very stable and
statistically consistent. The case $N=6$ reduces the systematic error
in $\alpha_s$ by factor $1.8$ and increases the value of $\alpha_s$,
while increasing the statistical error on the 30\%.

After the cuts have been implemented 
%(in this Section below 
(hereafter we use the set 
$N=6$),
we have 452 points in the analysis.
Fitting them in agreement with the same procedure considered in the previous 
Section,
we obtain the following results:
\bea
\as(M_Z^2) &=& 0.1153 \pm 0.0013 ~\mbox{(stat)} 
\pm 0.0022 ~\mbox{(syst)} \pm 0.0012 ~\mbox{(norm)},
\nonumber\\
&=& 0.1153 \pm 0.0028 ~\mbox{(total experimental error)} 
\nonumber
\eea
where
hereafter the symbol 
%``stat'', ``syst'' and 
``norm'' marks the 
%statistical error, systematic one and the 
error of normalization of experimental data.
The total experimental error is squared root of sum of squares of 
statistical error, systematic one and error of normalization.

\vspace{-0.2cm}
\subsection { SLAC, BCDMS, NMC and BFP data }

After these cuts have been incorporated (with $N=6$) for BCDMS data, 
the full set of combine data is 797 points.

%\vspace{-0.5cm}
To verify the range of applicability of perturbative QCD,
we analyze firstly the data without a contribution of twist-four terms,
i.e. when $F_2 = F_2^{pQCD}$. We do several fits using the cut 
$Q^2 \geq Q^2_{cut}$ and increase the value $Q^2_{cut}$ step by step.
We observe  good agreement of the fits with the data when 
$Q^2_{cut} \geq 10$ GeV$^2$ (see the Fig. 2).

Later we add the twist-four corrections and fit the data with the
usual cut $Q^2 \geq 1$ GeV$^2$.
We have find very good agreement with the data. Moreover 
the predictions for $\asMZ$ in both above procedures 
are very similar (see the 
%Table 6 and 
Fig. 2).

\begin{figure}[tb]
\begin{center}
\psfig{figure=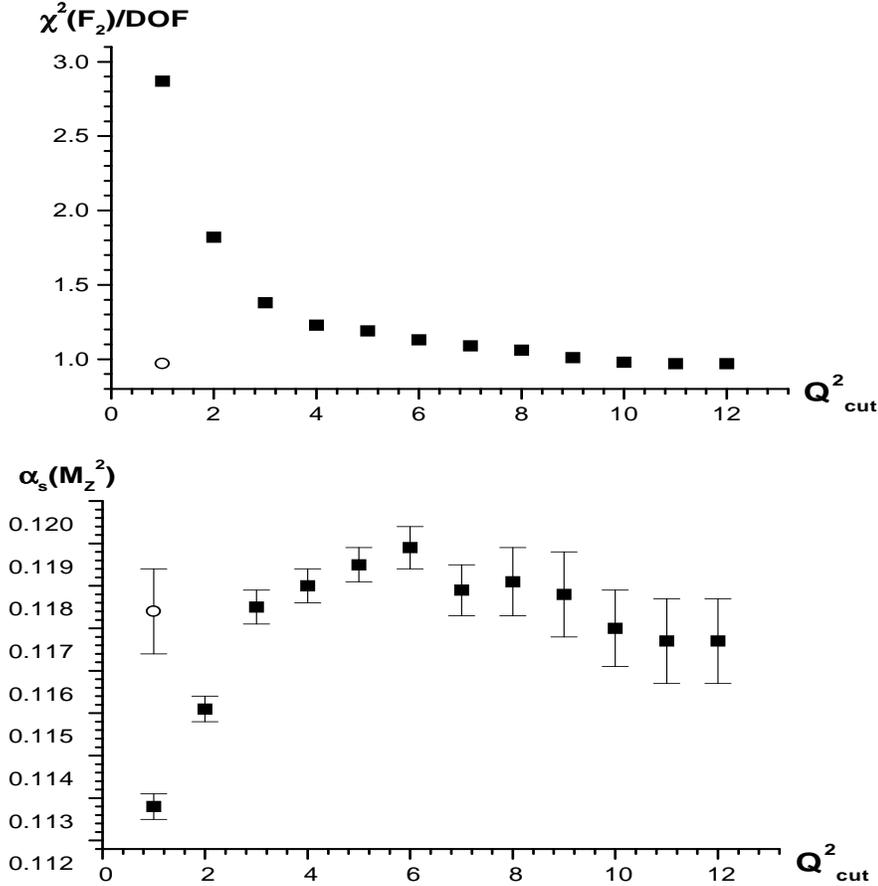,width=12cm,height=13cm}
\end{center}
\vskip -1cm
\caption{
The values of $\asMZ$ and $\chi^2$ at different $Q^2$-values of data cutes.
%in the fits based on nonsinglet evolution.
The black 
points show the 
analyses of data without  twist-four contributions.
The white point corresponds to the case where twist-four contributions 
were added.  Only statistical errors are shown.
}
\label{Non}
\end{figure}

So, the analysis of combine SLAC, NMC, BCDMS and BFP data 
are given the following results:
\begin{itemize}
\item  
When twist-four corrections are not included and the cut of $Q^2$ is 
10 GeV$^2$ at the free normalization 
\bea 
\chi^2\mbox{/DOF}~=~0.98~~~\mbox{ and }~~~
\as(M_Z^2) &=& 0.1170 \pm 0.0009 ~\mbox{(stat)}
\label{fu1.1}
\eea
\item  
When twist-four corrections are included and the cut of $Q^2$ is 1 GeV$^2$ 
\bea 
\chi^2\mbox{/DOF}~=~0.97~~~\mbox{ and }~~~
\as(M_Z^2) &=& 0.1174 \pm 0.0010 ~\mbox{(stat)}
\label{fu1.2}
\eea
\end{itemize} 

\vskip -0.3cm
Thus, as it follows from nonsinglet fits of experimental data, 
perturbative QCD
works rather well at $Q^2 \geq 10$ GeV$^2$.

\vspace{-0.5cm}
\section{ Summary }

We
have demonstrated several steps of our study \cite{KriKo}
of the $Q^2$-evolution of DIS structure function $F_2$ fitting all
modern experimental data 
at Bjorken variable $x$ values: $x \geq 10^{-2}$. 

%{\bf 1.} 
From the fits we have obtained the value of the normalization 
$\asMZ$
of QCD coupling constant. First of all, we have reanalyzed the BCDMS data 
cutting the range with large systematic errors. As it is possible to see
in 
the Fig. 1, 
the value of $\asMZ$ rises strongly when
the cuts of systematics were incorporated. In another side, 
the value of $\asMZ$ does not dependent on the concrete type of the
cut within 
%in the range of 
modern statistical errors.

We have found that at $Q^2 \geq 10 \div 15$ GeV$^2$ 
the formulae of pure perturbative
QCD (i.e. twist-two approximation together with target mass corrections)
are in good agreement with all data. 
\footnote{We note that at small $x$ values, the perturbative QCD
works well starting with $Q^2 = 1.5 \div 2$ GeV$^2$
and higher twist corrections are important only at very low $Q^2$:
$Q^2 \sim 0.5$ GeV$^2$ (see \cite{Q2evo,HT} and references therein).
As it is was observed in \cite{DoShi,bfklp} (see also discussions in
\cite{Q2evo,HT,BoAnd}) the good agreement between perturbative QCD and
experiment seems connect with large effective argument of coupling
constant at low $x$ range.}
 The 
results for  $\asMZ$ are very similar (see \cite{KriKo}) for the 
both types of analyses: ones, based on
nonsinglet evolution, and ones, based on combined singlet and 
nonsinglet evolution.
%ones and singlet ones. 
They have the following form:
\begin{itemize}
%
%\vskip -0.3cm
\item  from fits, based on nonsinglet evolution:
%\vskip -0.3cm
\bea 
\as(M_Z^2) &=& 0.1170 \pm 0.0009 ~\mbox{(stat)}
\pm 0.0019 ~\mbox{(syst)} \pm 0.0010 ~\mbox{(norm)}, \label{re1n} 
\eea
\item from fits, based on combined singlet and 
nonsinglet evolution:
\bea
\as(M_Z^2) &=& 0.1180 \pm 0.0013 ~\mbox{(stat)}
\pm 0.0021 ~\mbox{(syst)} \pm 0.0009 ~\mbox{(norm)}, 
\label{re1s}
\eea
\end{itemize}
\vskip -0.3cm

When we have added twist-four corrections, we have very good agreement
between QCD (i.e. first two coefficients of Wilson expansion)
and data starting already with $Q^2 = 1$ GeV$^2$, where the Wilson
expansion should begin
%start 
to be applicable.
The results for  $\asMZ$ coincide for the both types of analyses:
%nonsinglet ones and singlet ones. They have the following form:
ones, based on
nonsinglet evolution, and ones, based on combined singlet and 
nonsinglet evolution.
%ones and singlet ones. 
They have the following form:
\begin{itemize}
\item  from fits, based on nonsinglet evolution:
\bea 
\as(M_Z^2) &=& 0.1174 \pm 0.0007 ~\mbox{(stat)}
\pm 0.0019 ~\mbox{(syst)} \pm 0.0010 ~\mbox{(norm)}, \label{re2n} 
\eea
\item from fits, based on combined singlet and 
nonsinglet evolution:
\bea
\as(M_Z^2) &=& 0.1177 \pm 0.0007 ~\mbox{(stat)}
\pm 0.0021 ~\mbox{(syst)} \pm 0.0009 ~\mbox{(norm)}, 
\label{re2s}
\eea
\end{itemize}
\vskip -0.3cm

Thus, there is very good agreement (see Eqs. (\ref{re1n}), (\ref{re1s}),
(\ref{re2n}) and (\ref{re2s}))
between results based on pure perturbative QCD at quite large $Q^2$ values
(i.e. at $Q^2 \geq 10 \div 15$ GeV$^2$) and the results based on 
%fits with using of 
first two twist terms
%coefficients 
of Wilson expansion (at $Q^2 \geq 1$ GeV$^2$, 
where the Wilson expansion should  be applicable).

We would like to note that we have good agreement also with the analysis 
\cite{H1BCDMS} of
combined H1 and BCDMS data, which has been given by H1 Collaboration very 
recently. 
Our results for $\as(M_Z^2)$ are in good agreement also with 
the average value for coupling constant,
%for $\asMZ$, 
presented in the recent studies (see \cite{Al2000,NeVo,SaYnd,LEP}
and references therein) and in
famous Altarelli and Bethke reviews \cite{Breview}.

\vskip 0.2cm
{\bf Acknowledgments.}~~Authors
 would like to express their sincerely thanks to the Organizing
  Committee of the XVIth International Workshop ``High Energy Physics
and Quantum Field Theory'' for the kind invitation,  
the financial support
 at  such remarkable Conferences, and 
 for fruitful discussions.
A.V.K. was supported in part by Alexander von Humboldt
fellowship and INTAS  grant N366.

\end{document}